\documentclass{aastex}
\usepackage{spr-astr-addons}
\usepackage{url}

\RequirePackage{color}

\begin{document}

\title{Continuous Monitoring of Comet Holmes from Before the 2007 Outburst}
\slugcomment{Not to appear in Nonlearned J., 45.}
\shorttitle{Continuous Monitoring of Comet Holmes 2007 Outburst}
\shortauthors{E. E. El-Houssieny,  et al.}

\author{Ehab E. El-Houssieny\altaffilmark{1}} \and \author{Robert J. Nemiroff\altaffilmark{1}}
\affil{Michigan Technological University, Department of Physics,
1400 Townsend Drive, Houghton, MI 49931, USA} \and
\author{Timothy E. Pickering\altaffilmark{2}}
\affil{MMT Observatory, University of Arizona, Tucson, AZ 85721,
USA}

\begin{abstract}
The outburst and subsequent brightness evolution of Comet
17P/Holmes has been observed using the MMT Observatory's All-Sky
Camera \citep{Pic06} on Mt. Hopkins near Tucson, Arizona, USA. The
comet was picked up at the limiting visual magnitude of 5.5 on
October 24.38 and tracked by the camera continuously until sunrise
four hours later. During this time the comet brightened to visual
magnitude 3.5. Comet Holmes was next observed just after sunset on
October 25.23 at visual magnitude 2.5 where it remained
approximately constant over the next three days. The comet then
began to dim slowly and was followed into the early months of 2008
with periods of dense time coverage.
\end{abstract}


\section{Introduction}

Comets have been noted in the sky for almost as long as history
has been recorded.  A verifiable brightness variation for any
bright comet is rare, however, since human observers are hard to
calibrate, photographic magnitudes can be unreliable for bright
objects, and because telescopes with modern CCDs typically have
fields of view too small to contain a comet that has reached
naked-eye visibility.  Even for comets monitored by CCD, unusual
cometary events are rarely recorded, since observers are typically
alerted to look only after such an event has occurred.

In recent years, however, a class of all-sky cameras has begun to
be used routinely in astronomy, primarily to assess sky conditions
\citep{Nem99, Pic06, Sha05a, Sha05b}. These cameras typically
utilize fisheye lenses with fields of view in excess of 150
degrees and can thus capture images of a comet of almost any size.
Furthermore, these cameras typically operate every clear night,
and so are likely to be operating during an unusual cometary
outburst.

In 2007 October, periodic comet Holmes underwent an unusual
outburst which increased its brightness from an apparent visual
magnitude of about 17 to near 3.  This paper reports details from
an all-sky camera that captured this cometary outburst and
monitored the brightness of the comet continuously for the next
few hours and for many nights over the next three months.

\section{A Brief History of Comet Holmes}

Comet Holmes was discovered by Edwin Holmes (London, England) on
November 6, 1892 within the Andromeda Galaxy (M31). The comet
brightened from dimmer than visual magnitude 17 to about visual
magnitude 2.8 over about 42 hours, creating a coma about 5 arc
minutes diameter (Whipple 1984).  The comet's discovery was
confirmed by Edward Walter of the Royal Observatory in Greenwich,
England \citep{Whi84, bob43}.  Unexpectedly, comet Holmes
underwent a second outburst only few months later, on January 16,
1893.  In the second outburst, the comet brightened to visual
magnitude 8 and exhibited a coma of 41 arc seconds in diameter.
The comet steadily exhibited a larger coma until late the next
night and then steadily faded after the next outburst. The last
observation before fading from visibility was made by H. C. Wilson
of Goodsell Observatory in Northfield, Minnesota on April 4, 1893
\citep{bob43, Zwi12}. The comet was lost after 1906 until being
re-acquired on July 16, 1964 by Elizabeth Roemer of the Naval
Observatory in Arizona, USA (Whipple 1984).

Several attempts have been made to determine the comet's orbital
elements. The first orbit determination was calculated by H. C.
Kreutz using three positions measured on November 9, 10, and 11,
1892 (Zwiers 1912).  Kreutz discussed the difficulty of
calculating comet Holmes' orbit and introduced four potential
parabolic orbits satisfying three observations with perihelion
passage time ranging from February 28 - June 7, 1892 and an
orbital period of 6.9 years. During the next few weeks, several
more attempts were made to more precisely determine comet Holmes'
orbit.  These attempts also derived rather different orbital
elements with perihelion passage estimates ranging from February
28 to August 16, 1892.

\section{The Outburst and the Brightness Variation of Comet Holmes in
2007-2008}

\begin{figure}[t]
\includegraphics[scale=0.33, angle=0]{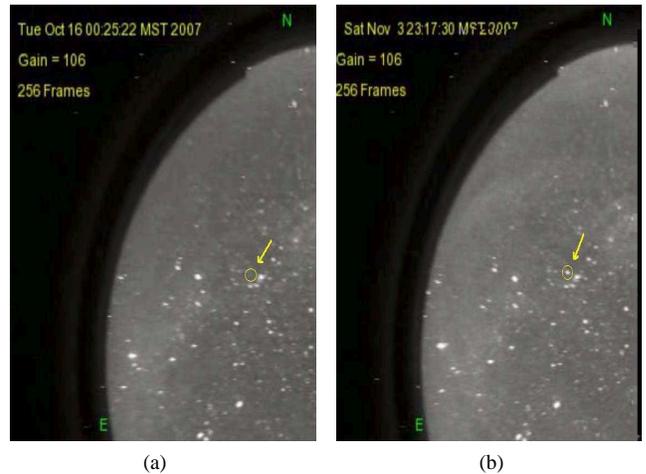}
\caption{Two images extracted from MMTO All Sky Camera image
archive.  Comet Holmes is barely visible on the image on the left,
taken on October 16 when it was fainter than 17th magnitude, while
only 18 nights later, at the same sidereal time, the comet is
easily visible on the image on the right at a visual magnitude of
2.8.} \label{Ehab_fig1}
\end{figure}

On October 24, 2007, comet Holmes underwent an outburst similar to
its first outburst.  During the early hours of October 24, 2007,
the comet became much brighter, increasing its brightness from a
visual magnitude of about 18 to 2.5 over less than two days. Comet
Holmes became the third brightest object in the constellation of
Perseus (see Figure~\ref{Ehab_fig1}-b) and was visible to the
unaided eyes of even casual observers.

Although hampered by moonlight, the MMT All-Sky Camera was able to
capture comet Holmes on the night of its sudden brightening.
Through the course of its normal operation, the All-Sky Camera was
then able to follow the evolution of comet Holmes for several
months after its outburst. On a dark night, the system defaults to
an 8.533 second exposure time which results in a limiting
magnitude of about 5.5 in V and 6 in R. Sensitivity is decreased
by moonlight, however, due to glare, reduced gain, and reduced
exposure time.

\begin{figure*}[!ht]
\centering
\includegraphics[scale=0.50, angle=0]{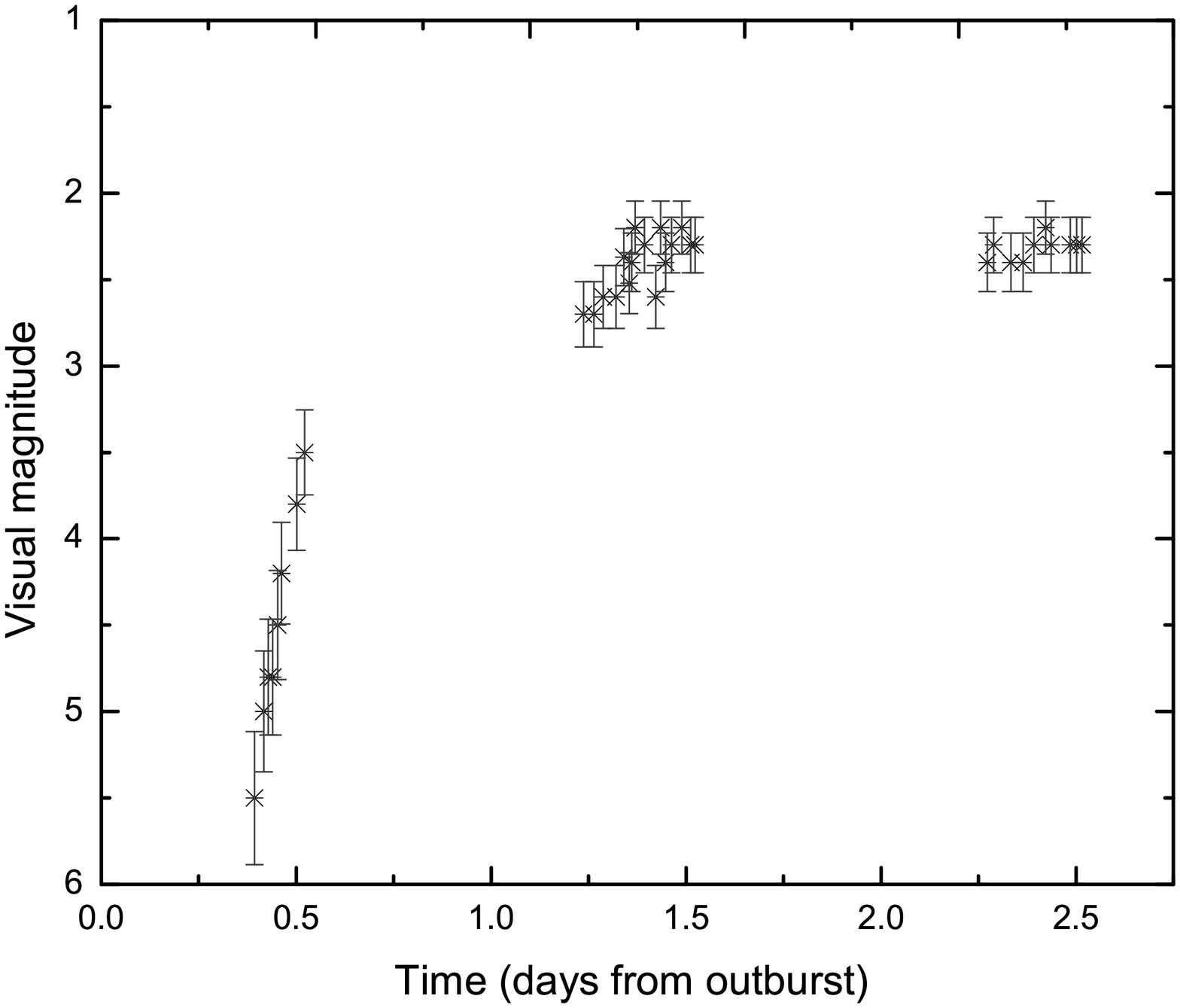} \caption{The Comet Holmes' brightness variation from October 24-26, 2007.}\label{Ehab_fig2}
\includegraphics[scale=0.50, angle=0]{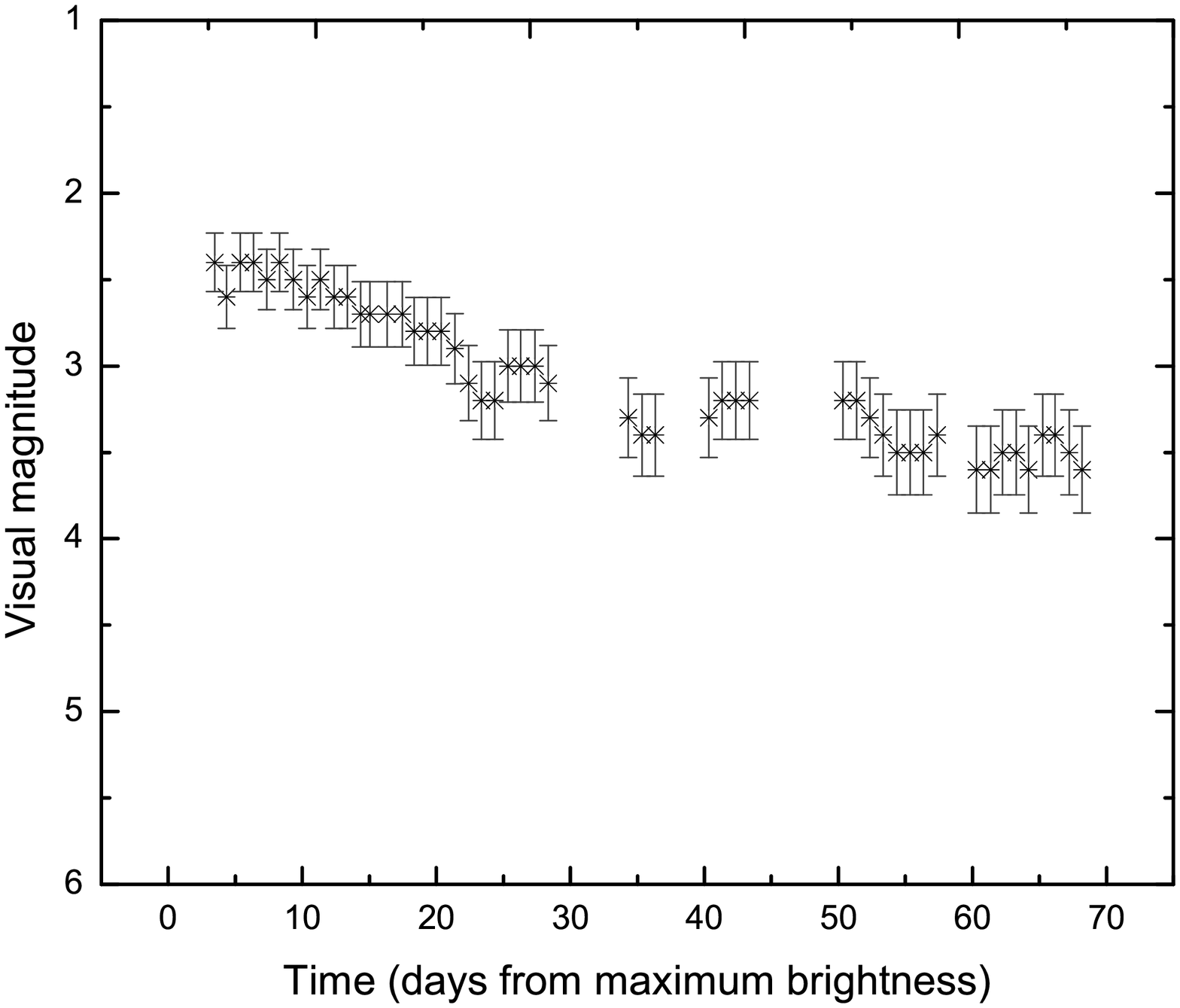}
\caption{The Comet Holmes' brightness variation from October 27,
2007 - January 1, 2008.} \label{Ehab_fig3}
\end{figure*}

It was necessary to divide comet Holmes' brightness variation into
two graphs according to our continuous observations of the comet
from the time it became visible on MMTO All Sky Camera images over
the next three months. Figure~\ref{Ehab_fig2} shows a plot of the
visual magnitude of comet Holmes over the first three days where
the comet exhibited extreme magnitude change from below visibility
to magnitude 2.5 in less than 24 hours. Data points are plotted
for every hour by averaging over each 10-second exposure. While
Figure~\ref{Ehab_fig3} shows a plot of the comet magnitude over
the next three months where the comet exhibits steady fading
phase. Data points are plotted for every notable magnitude change,
nearly for every 24 hours. Magnitude estimates in both graphs were
made by comparison to stars of cataloged magnitudes (\cite{Hen}).

Figure~\ref{Ehab_fig2} shows that comet Holmes started its
outburst at the early hours of October 24, 2007, in particular
about 9.5 Universal Time (UT) and it was of magnitude 5.5 to be
visible to unaided eyes. In less than 24 hours, comet Holmes
surprisingly brightened to magnitude 3.5 that is enough to be seen
in full-moon night and town light pollution. In the next night,
October 25, 2007, at 06:53 am UT, the comet reached magnitude 2.5.
This is followed by complete constancy at maximum light sustained
for 2 days after the comet outburst.

Figure~\ref{Ehab_fig3} shows that comet Holmes remained visible
over the next three months. Night to night, from October 27, 2007
- January 1, 2008, comet Holmes is slowly and steadily fading
through periods of about three days long.



\section{Quantifying Comet Holmes' Brightness Variation Curve}

\begin{figure*}[!ht]
\centering
\includegraphics[scale=0.55, angle=0.0]{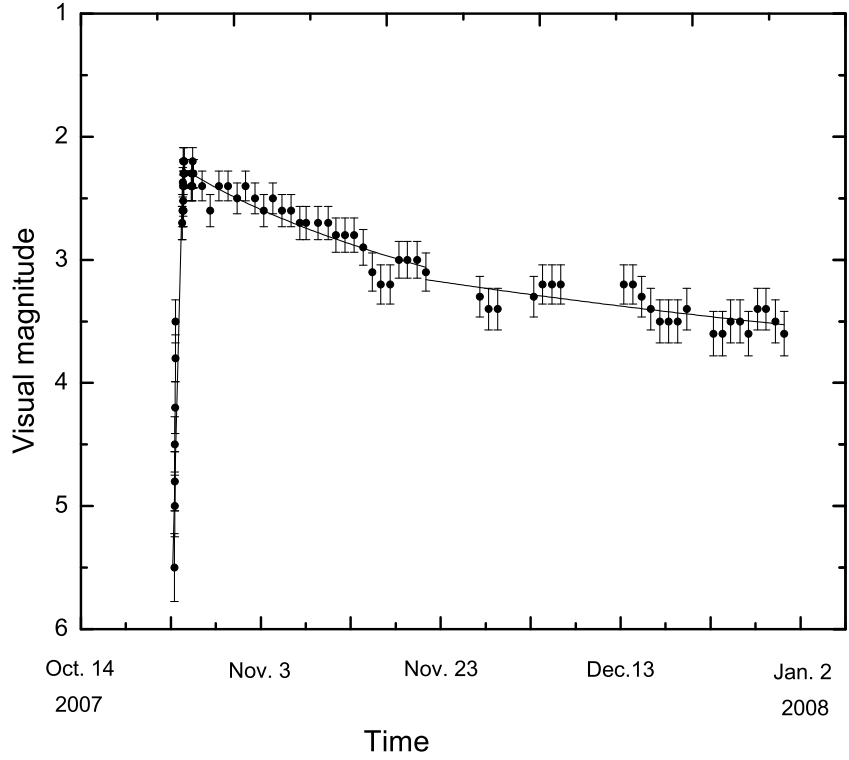}
\caption{Comet Holmes brightness variation curve according to
magnitude estimates is drawn by dots and the solid line represents
the secular variation of the brightness fitted by the linear least
square method.} \label{Ehab_fig4}
\end{figure*}

For explanatory and future predictive value of the photometric
parameters, an attempt was made to quantify the evolution of the
brightness of comet Holmes as a function of the heliocentric and
geocentric distances. It is well known that the dust jets of
active comets may take tens of days to dissipate by solar
radiation pressure into a comet's tail, especially at large
heliocentric distances, and there might be an unknown contribution
of a comet's tail at small phase angles because it lies on the
line-of-sight of the comet (Gonzales 1969; Ip, W.-H. 1979). The
2007 outburst of comet Holmes occurred at relatively large
heliocentric distance and almost zero phase angle. During the
first month from the outburst the comet dimmed rapidly with
dimming rate of 0.02 magnitude/day, while after November 23, 2007,
the comet experienced a slow and steady dimming phase with dimming
rate of 0.0075 magnitude/day, see Figure \ref{Ehab_fig4}. This
slow and steady dimming phase might be interpreted as due to the
increase in the heliocentric distance because the short term dust
dissipation constitutes a small fraction of the total brightness
variation. Such reasoning along with precise total brightness
estimates taken by single instrument with small aperture, such as
MMT all-sky camera (aperture $<1$ cm), may allow the use of the
approximated equation of the total magnitude of a comet with coma,
represented by

$$
m = H_0 + 5\, log(\Delta) + 2.5 \, n \, log (r),
$$
or by rearranging the terms to be in the form,
\begin{equation}
\label{eq1} \hspace{1.7 cm} m - 5\, log(\Delta) = H_0 + 2.5 \, n
\, log (r),
\end{equation}
where $H_0$ is the absolute magnitude of the comet, $\Delta$ and
$r$ are the geocentric and the heliocentric distances (AU)
respectively, and $n$ is the exponent of the heliocentric distance
power-law (Fernandez, 2005).

\begin{figure*}[!ht]
\centering
\includegraphics[scale=0.6, angle=0.0]{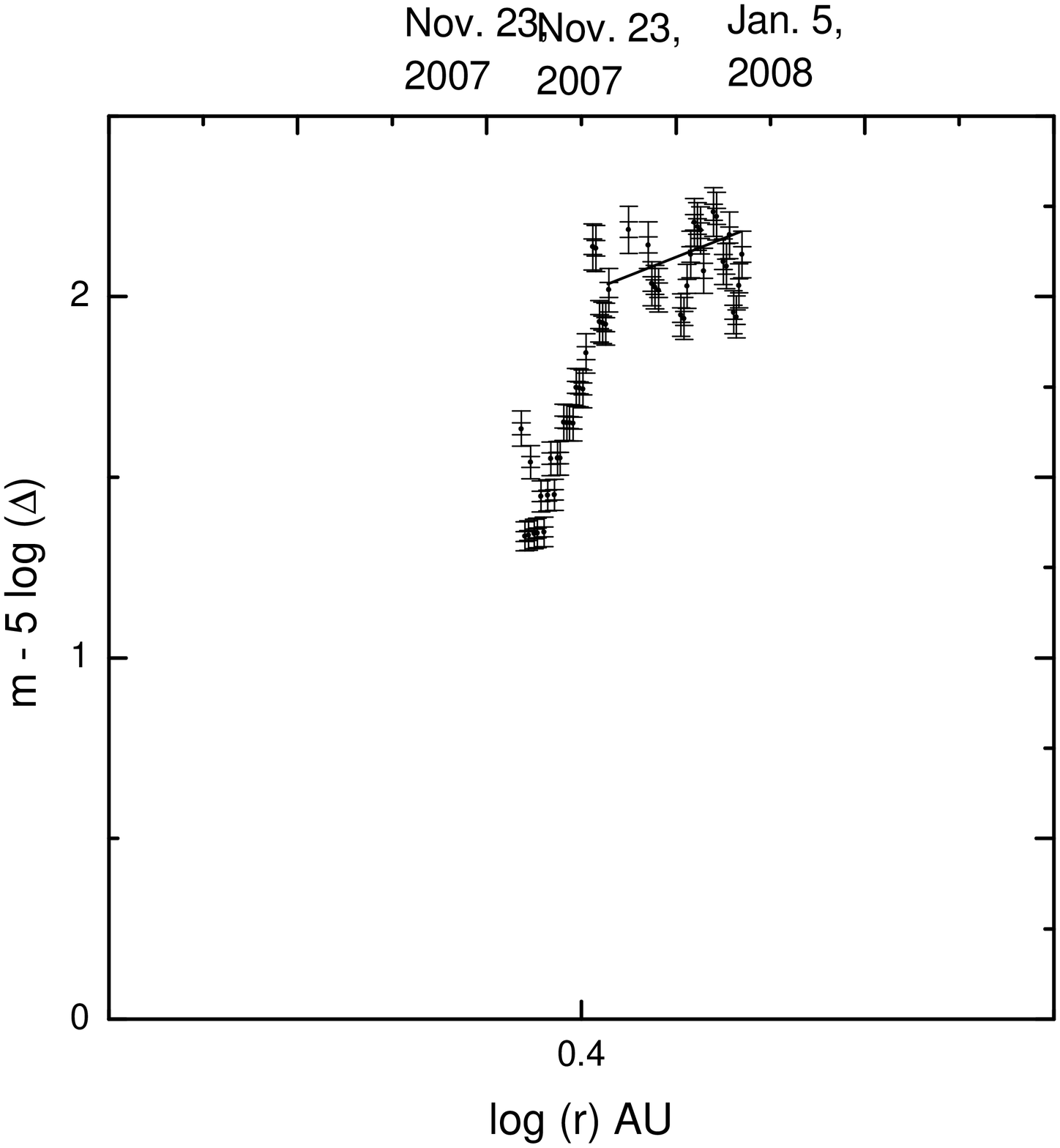}
\caption{The total brightness variation of comet 17P/Holmes with
the linear fit representing the solution to equation
(\ref{eq1}).}\label{Ehab_fig5}
\end{figure*}

To investigate the variation of the photometric parameters $H_0$
and $n$ of comet Holmes, we have plotted the linear equation
(\ref{eq1}) and used linear fitting in an attempt to minimize the
coma and the outburst consequences. Figure \ref{Ehab_fig5} shows
that there is a phase transition had occurred on about November
23, 2007, that might be interpreted as the beginning of the normal
phase of the brightness variation of a comet, in which the
brightness variation depends on the geocentric and the
heliocentric distances as well as the phase angle according to
equation (\ref{eq1}). The resulted photometric parameters are
summarized in Table (\ref{table-1}).

\begin{table*}[hb]
\caption{The photometric parameters of comet 17P/Holmes after its
2007 outburst.}
\begin{tabular}{c c c c}
\hline\hline
Time range& Range of r (AU) & $H_0 + 2.5\, n\, log(r)$  \\
\hline
Oct. 24 - Nov. 23, 2007 & 2.436 - 2571 & ---  \\
Nov. 23, 2007 - Jan. 1, 2008 & 2.571 - 2.717 & 1.13($\pm 0.76387$)+\,2.25\,($\pm 1.81734$)\,$log(r)$ \\
\hline
\end{tabular}
\label{table-1}
\end{table*}


\section{Summary and Conclusions}


Comet outbursts were discovered many decades ago (Bobrovnikoff
1943; Zwiers 1912), though a complete understanding as to why some
comets under an outburst has remained elusive. However, comet
brightness variation curves provide important information about
comet flares (see, for example, Churyumov and Filoneko 1993). In
an attempt to provide a clearer picture about this unusual
cometary flare, our paper summarizes and analyzes the results of
observations of comet Holmes' flare that began in October 2007.
These observations were taken with the MMT All Sky Camera over
three months from the beginning of its outburst. From the time
comet Holmes became visible to the MMTO All Sky Camera, it
brightened from visual magnitude 5.5 to magnitude 2.5 in less than
29 hours, and then followed by constancy at maximum brightness
over the next two days. After reaching its peak, comet Holmes
steadily faded through periods of about 3 days long to fainter
magnitudes over the next three months to reach magnitude 3.7 on
January 1, 2008. There is a phase transition had occurred on about
November 23, 2007, in which the dimming rate was much smaller
(0.0075 magnitude/day) and the photometric parameters was
dramatically changed.

To the best of our knowledge, no comet has ever been monitored by
a single dedicated instrument for so long a period of time
previously. The MMT All Sky Camera has proved useful in that it
can continuously monitor the variation of a comet's total visual
brightness. It is our hope that the comet Holmes' brightness
variation curves presented here will provide useful constraints
for future comet outburst models.


\acknowledgments

We are grateful to the anonymous referees for valuable comments
and suggestions that greatly improved our manuscript in all
aspects.

\nocite{*}
\bibliographystyle{spr-mp-nameyear-cnd}
\bibliography{biblio-u1}

\end{document}